\documentclass[aps,prl,a4paper,twocolumn,longbibliography,superscriptaddress]{revtex4-2}
\usepackage[utf8]{inputenc}
\usepackage{amsmath,amsfonts,amssymb}
\usepackage{mathtools}
\usepackage{graphicx,xcolor}
\usepackage[space]{grffile}   
\usepackage[normalem]{ulem}     

\newcommand{\beq}{\begin{equation}}
\newcommand{\eeq}{\end{equation}}
\newcommand{\bei}{\begin{itemize}}
\newcommand{\eei}{\end{itemize}}
\newcommand{\ben}{\begin{enumerate}}
\newcommand{\een}{\end{enumerate}}

\newcommand{\be}{{\mathbf e}}

\usepackage{hyperref}
\hypersetup{
   pdfpagemode=UseNone, 
   pdfstartpage=1,
   pdfmenubar=true,
   pdftoolbar=true,
   colorlinks=true,
   linkcolor=blue,
   citecolor=blue,
   urlcolor=blue,
   bookmarksopen=false
 }

\definecolor{darkblue}{rgb}{0.,0.24,0.51}
\definecolor{britishracinggreen}{rgb}{0.0, 0.26, 0.15}
\definecolor{darkgreen}{rgb}{0,0.60,.2}

\def\be{\begin{equation}}
\def\ee{\end{equation}}

\def\rf#1{(\ref{#1})}

\begin{document}
\title{Fermi polarons with the finite-range fermion-impurity interactions}
\author{Nikolay Yegovtsev}
\affiliation{Department of Physics and Astronomy and IQ Initiative, University of Pittsburgh, Pittsburgh, Pennsylvania 15260, USA}

\date{\today}
	
\begin{abstract}
We study a problem of an infinitely heavy impurity introduced into a polarized Fermi gas, which can be solved exactly with the help of Fumi's theorem. We consider the regime of finite-range fermion-impurity interactions beyond the standard $s$-wave scattering regime and investigate how this affects the energy of the polaron. We show how one can account for the effective range effects as well as the contribution from higher angular momentum channels, which are important for the study of ionic and Rydberg polarons. Our findings have relevance for atomic gas mixtures with a large mass imbalance and for the impurities trapped inside the optical tweezers.
\end{abstract}
\maketitle

The problem of a single impurity interacting with particles in its host environment has attracted significant attention from the atomic physics community in recent years \cite{massignan2025polarons, Grusdt_2025}. This interest comes from the possibility of engineering the bath-particle interactions with great precision and control, which allows for experimental probing of such systems in the regime of strong interactions and eventually guiding us in the quest for understanding the physics of strongly correlated phenomena. The problem of a mobile impurity introduced into a polarized gas of fermions is called the Fermi polaron problem. Originally, people were interested in the dilute limit $k_Fr_0\ll1$, where $r_0$ is the characteristic range of the microscopic potential, and the only relevant parameter characterizing the fermion-impurity scattering is the $s$-wave scattering length $a$. Scattering length can be tuned experimentally using a magnetic Feshbach resonance \cite{chin2010feshbach}, and the regime of strong interactions corresponds to $a\to\infty$. Despite its simplicity, this problem cannot be solved exactly; however, Chevy proposed a simple and elegant variational solution \cite{chevy2006universal}, which turned out to be in remarkable agreement with the experiment \cite{yan2019boiling}. The Fermi polaron problem in 3D, where the impurity atom has the same mass $M$ as the mass of the fermionic bath particles $m$, has been extensively studied in the literature \cite{massignan2025polarons}. One peculiar fact about this problem is that it becomes exactly solvable in the special limit $M\to \infty$, because in this limit the impurity acts as a scattering center for all fermions \cite{mahan2013many, combescot2007normal}. If the system is placed in a large box, then the expression for the energy of the polaron is provided by Fumi's theorem, so one can extract the properties of the polaron by studying the corresponding scattering phase-shifts $\delta_l(k)$ and possible bound states $\varepsilon_n$ \cite{mahan2013many}:
\begin{equation}
\label{eq:exactenergy}
E = -\frac{\hbar^2}{m\pi}\sum_{l}(2l+1)\int_0^{k_F}dk\,k\delta_l(k) +\sum_n\varepsilon_n.    
\end{equation}
What is surprising is that in the universal regime where $k_Fr_0\ll1$ and $a\to\infty$, the answer based on Fumi's theorem is only 18 percent off from the corresponding quantity for the $M=m$ case \cite{combescot2007normal}. One would expect that if the mass of the impurity is made much larger than the mass of the fermions, the predictions based on Fumi's theorem would become not only qualitative, but also quantitative. Luckily, this theoretically idealized situation can be achieved experimentally by working with the atomic gas mixtures with large mass imbalance \cite{PhysRevLett.106.205304,PhysRevA.84.011606,PhysRevLett.95.170408,PhysRevLett.119.233401}, so that $M\gg m$ is satisfied, or trapping the impurity atom into a species-selective optical tweezer which is transparent for the rest of the fermions. 

What is special about Eq.~\rf{eq:exactenergy} is that for a given fermion-impurity potential, it contains nonperturbative information about scattering in all angular momentum channels, so for a given value of $k_Fr_0$ we can deduce which physical effects are dominant. For example, starting from the regime $k_Fr_0\ll1$, one can extend the universal results for the short-ranged potentials down to $ k_Fr_0\lesssim 1$. At the same time, for the regime $k_Fr_0\gtrsim1$, which is experimentally relevant for the ionic and Rydberg polarons \cite{zipkes2010trapped, massignan2025polarons}, we can study the relevance of higher angular momentum channels. Since a simple qualitative and quantitative description of the mobile Fermi polarons in the spirit of the Chevy ansatz beyond the regime of $s$-wave scattering is currently missing \cite{massignan2025polarons}, we expect that the analysis based on Fumi's theorem can serve as a first step in that direction. We also note a recent development \cite{chen2025massgapdescriptionheavyimpurities} on the theory of mobile impurities that connects to the results of the static impurity of Eq.~\rf{eq:exactenergy} in the regime of $s$-wave scattering. This construction is also possible for finite-ranged potentials \cite{chen2025massgapdescriptionheavyimpurities}. Although Fumi's theorem has previously been discussed in the polaron literature, its applications have been confined to the regime of $s$-wave interactions $k_Fr_0\ll1$ \cite{zollner2011polarons, PhysRevB.101.195417, PhysRevA.102.023304, PhysRevResearch.4.013160, chen2025massgapdescriptionheavyimpurities}.  Finally, it is possible to derive an analog of Eq.~\rf{eq:exactenergy} in lower dimensions \cite{giraud2009highly, zollner2011polarons}, so the present analysis in 3D can be carried there as well.

In this Letter, we show that all essential properties of a heavy Fermi polaron can be extracted from Eq.~\rf{eq:exactenergy}. To make the analysis more transparent, we model the impurity potential by the attractive square well potential:
\begin{equation}
\label{eq:swpotential}
V = \begin{cases}
			-\frac{\gamma^2\hbar^2}{2mr_0^2} & \text{if $r\leq r_0$},\\
            0 & \text{otherwise}.
		 \end{cases}
\end{equation}
The expression for the phase shifts reads \cite{sakurai2020modern, flugge2012practical}:
\begin{equation}
\label{eq:swdeltalm}
\tan\delta_l(x) = \frac{xj_l'(x)-y\frac{j_l'(y)}{j_l(y)}j_l(x)}{xn_l'(x)-y\frac{j_l'(y)}{j_l(y)}n_l(x)}, 
\end{equation}
where $x=kr_0$ and $y=\kappa r_0$, $\kappa = \sqrt{k^2+\gamma^2/r_0^2}$,  and $j_l, n_l$ are spherical Bessel functions. The first unitary point for the $s$-wave scattering occurs when $\gamma=\pi/2$. This potential is commonly used in Monte Carlo studies \cite{pessoa2021finite}, and as we will show, the Fermi polaron problem in the square well can be solved analytically in various regimes, so this toy model can play the role of a simple benchmarking tool for more elaborate techniques. We observe that the theory has two dimensionless parameters: $k_Fr_0$ and $\gamma$, so the properties of the polaron will depend on the interplay between both.

\begin{figure}[t]
\centering
\includegraphics[width=\columnwidth]{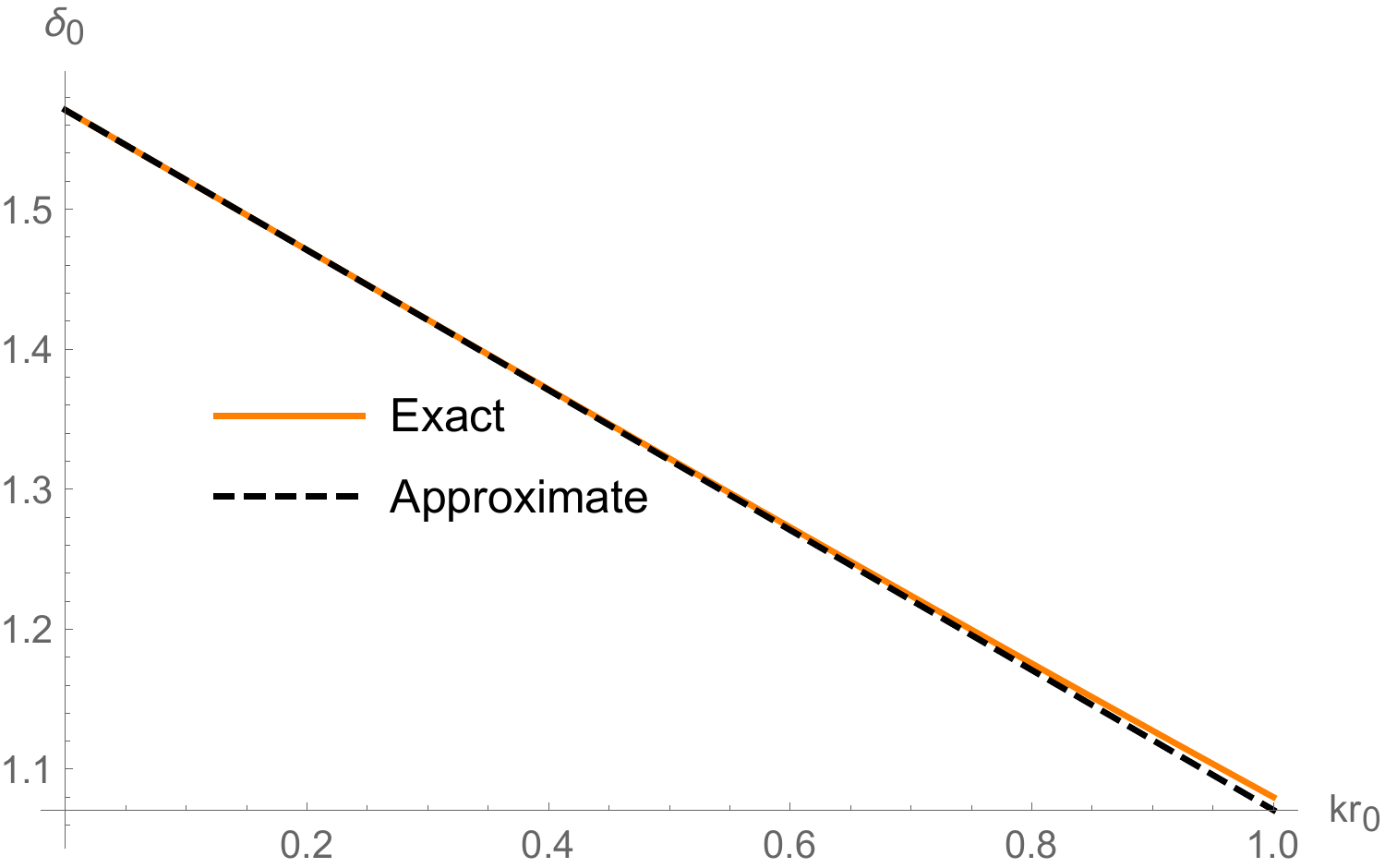}
\caption{
\label{fig:delta0} $l=0$ phase shift
$\delta_0$ for the square-well potential at the first unitary point $\gamma=\frac{\pi}{2}$. The exact expression in  given by Eq.~\rf{eq:swdeltalm} and the approximate one in Eq.~\rf{eq:delta0apprx}.
}
\end{figure}

\textit{General theory for $k_Fr_0\lesssim1$, $\gamma\leq\pi/2$} -- If the physical range of the potential $r_0$ is such that $k_Fr_0\ll1$, we can use the results from low energy scattering theory, which states that $\delta_l(k)\sim k^{2l+1}$ \cite{landau2013quantum}. Since $\delta$ is dimensionless, there should be some length-scale parameters multiplying powers of $k$. Typically, all such lengthscale parameters are of the order of $r_0$ \cite{hammer2010causality}, so this justifies keeping only the leading contribution $l=0$. In this regime $\delta_0(k)\sim-ak$, where $a$ is the $s$-wave scattering length. Plugging this into Eq.~\rf{eq:exactenergy}, and using the fact that the density of the polarized Fermi gas can be expressed as $n =k_F^3/(6\pi^2)$, we get:
\begin{equation}
\label{eq:mf}
E = \frac{2\pi \hbar^2 na}{m}.    
\end{equation}
This is the expected mean-field result. It can be further improved by noting that, according to the general theory of low-energy scattering, the scattering amplitude has the following regular expansion \cite{landau2013quantum}:
\begin{equation}
\label{eq:effrangeexp}
k\cot\delta_0(k) = -\frac{1}{a}+\frac{1}{2}r_\text{eff}k^2+\cdots,
\end{equation}
where $r_\text{eff}$ is the effective range, and we neglected higher-order contributions in the expansion. At the same time, we expect the polaron energy to be a continuous function of $\gamma$.  If we neglect the $r_\text{eff}$ term ($r_0k_F\ll1$), we can write $\delta_0 = \cot^{-1}\left(-1/ak\right)$ and obtain the expression for the polaron energy as a function of $a$ alone:
\begin{equation}
\label{eq:aonly}
E = -\frac{\hbar^2k_F^2}{2m}\left(\frac{a k_F-\left(a^2 k_F^2+1\right) \tan ^{-1}(a k_F)}{\pi a^2 k_F^2}\right).    
\end{equation}
This result is only applicable for $\gamma\leq\pi/2$, before the first unitary point is crossed. Note that at unitarity $a\to -\infty$ the above $M\to \infty$ results gives $E = -0.5E_F$. Let us focus on the unitary point $\gamma=\pi/2$ more closely. To comply with the Eq.~\rf{eq:effrangeexp}, $\delta_0(0) = \pi/2$, which implies that:
\begin{equation}
\label{eq:delta0apprx}
\delta_0(k) = \frac{\pi}{2}-\frac{1}{2}r_\text{eff}k+\cdots,    
\end{equation}
so when we plug the above expansion into Eq.~\rf{eq:exactenergy}, we get the effective range correction to the main result at unitarity:
\begin{equation}
\label{eq:Ewithreff}
E =     -\frac{1}{2}\left(\frac{\hbar^2k_F^2}{2m}\right) + \frac{\pi\hbar^2 nr_\text{eff}}{m}.
\end{equation}
On general grounds, we expect $r_\text{eff}\sim r_0$, so in the regime $k_Fr_0\ll1$ the contribution from the second term should be subleading. At the same time, if we increase the density of the Fermi gas or tune the size of the effective potential, as can be done in the case of the Rydberg impurity, we can start to probe the regime where $k_Fr_0\lesssim1$, where the effective range correction starts to become important. At this point, one has to start worrying about the contribution from the higher angular momentum channels. Let us show how a correction from the $l=1$ channel can be incorporated into the main result. The scattering amplitude of the $l=1$ channel has the following expansion \cite{gurarie2007resonantly}:
\begin{equation}
\label{eq:delta1effrangeexp}
k^3\cot\delta_1(k) = -\frac{1}{v}+\frac{1}{2}k_0k^2+\cdots,    
\end{equation}
where $v$ is the scattering volume and $1/k_0$ plays the role of the effective range. Since we are interested in the $s$-wave resonance, $v$ and $k_0$ should be finite, which implies the following form for $\delta_1$:
\begin{equation}
\label{eq:delta1approx}
\delta_1(k) = -vk^3  - \frac{1}{2}k_0v^2k^5+\cdots
\end{equation}
If we want to keep $k_0$ term, we would also need to include the $k^4$ term in the expansion of Eq.~\rf{eq:effrangeexp}, which is an effective expansion in powers of $k_Fr_0$. We need to consider all of those corrections if we want to push the theory close to $k_Fr_0\sim1$ from below, however as long as we are not too close to $k_Fr_0=1$, the expression in Eq.~\rf{eq:Ewithreff} gives an adequate answer as can be seen from the Fig.~\ref{fig:delta0} and noting that the leading contribution from $\delta_1$ will be suppressed relative to the effective range correction by one power of $k_Fr_0$. This shows that the original results for $k_Fr_0\ll1$ can be easily extended to the regime $k_Fr_0\lesssim1$ in a straightforward manner at the expense of adding a few more scattering parameters. The above construction was presented for the $s$-wave resonance, but a similar analysis can be performed around higher $l$ resonances. For this, we need to carefully include the presence of bound states in the lower $l$ channels. 

\textit{General theory for $k_Fr_0\lesssim1$, $\gamma\geq\pi/2$} -- Once we are past the unitary point, an impurity potential can support a bound state. For a shallow bound state, its energy is given by \cite{sakurai2020modern}:
$$\varepsilon = -\frac{\hbar^2}{2ma^2}.    $$
Note that one has to exercise caution when applying Eq.~\rf{eq:exactenergy}, since in the scattering theory phase shifts are defined modulo $\pi n$. In particular, once we are past the first unitary point, the result of Eq.~\rf{eq:aonly} will predict positive energy for the ground state, which is not physical. To address this problem, the phase shifts in Eq.~\rf{eq:exactenergy} have to be defined, so that they remain continuous as we cross the unitary points. In our case, this can be done by adding $\pi\Theta(a)$ term to the phase shift, where $\Theta(x)$ is the Heaviside step function. Including the contribution from the bound state as well, we finally arrive at:
\begin{equation}
\label{eq:aonly1}
E = -\frac{\hbar^2k_F^2}{2m}\left(\frac{a k_F+\left(a^2 k_F^2+1\right) \tan ^{-1}(\frac{1}{a k_F})}{\pi a^2 k_F^2}\right),    
\end{equation}    
which originally appeared in \cite{combescot2007normal}. In deriving this result for the ground state, we had to impose the continuity of the phase shifts as we crossed the unitary point. While the result of Eq.~\rf{eq:aonly} cannot be used to describe the ground state of the system for $a>0$, it is valid for excited states. Indeed, the repulsive polaron branch exists for $a>0$ and away from the resonance, it approaches the mean-field result of Eq.~\rf{eq:mf}, so at unitarity it predicts $E = 0.5E_F$ \cite{massignan2011repulsive}. The above analysis is applicable in the regime $k_Fr_0\ll1$, so as we start to increase it, we have to use the exact expression for the phase shifts and exact numerical values of the bound-state energy to accurately describe the polaron; however, the general procedure outlined above will hold true both for attractive and repulsive polaron branches beyond the regime of $s$-wave interactions. If the impurity potential supports multiple bound states, then there is a possibility for many excited state branches. Recently, a similar phenomenon has been discussed in the context of the Bose polaron problem in 3d \cite{Yegovtsev2022, mostaan2023unifiedtheorystrongcoupling}. In the present case, we expect that some of such branches can be constructed by using the phase shift branch that is continuously connected to the mean field result of Eq.~\rf{eq:mf}, similar to the case of the repulsive polaron.

\begin{figure}[t]
    \centering
    \includegraphics[width=\columnwidth]{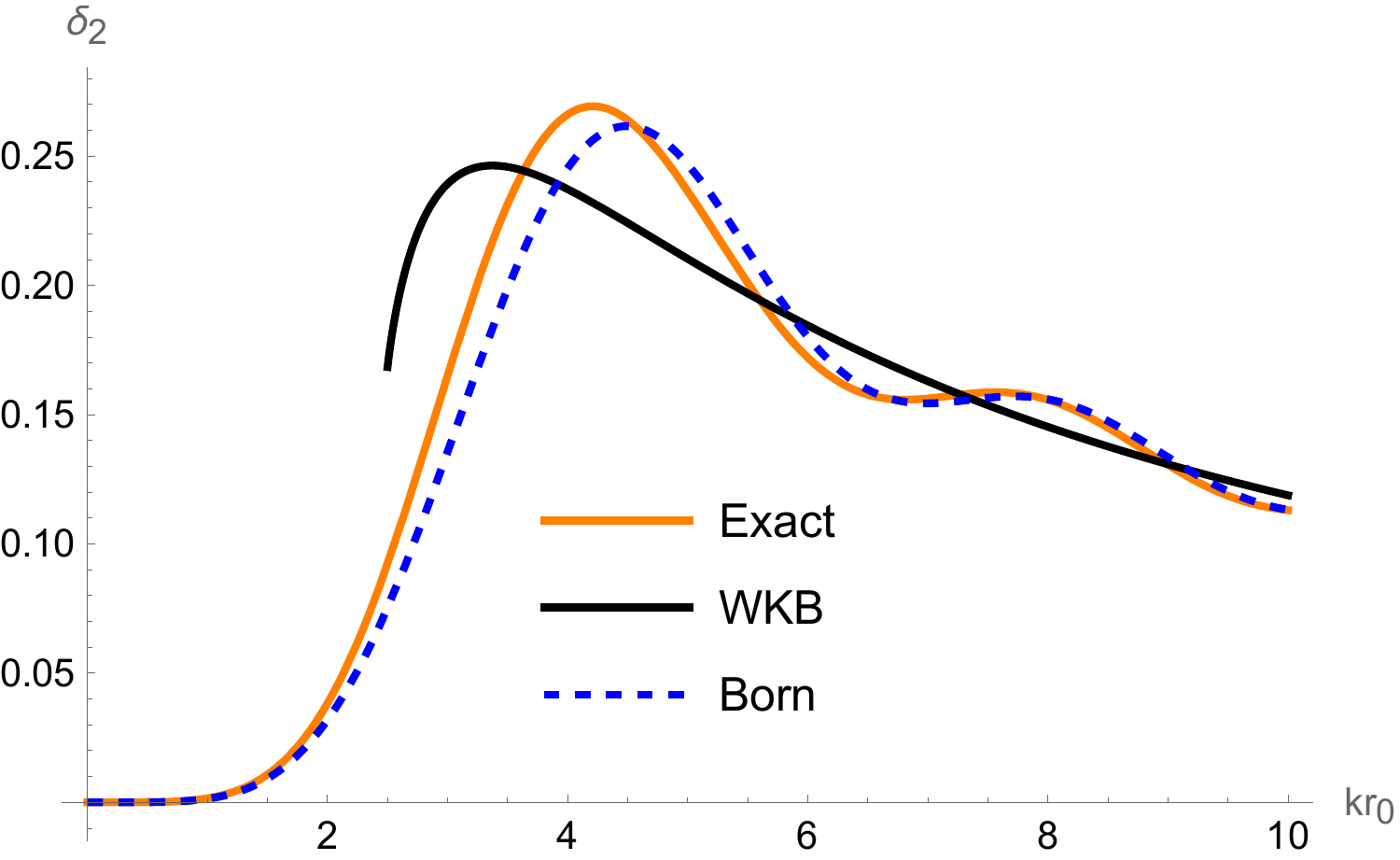}
    \caption{$l=2$ phase shift $\delta_{2}$ for the square-well potential at the first unitary point $\gamma=\frac{\pi}{2}$. The exact expression is given by Eq.~\rf{eq:swdeltalm}, WKB result by Eq.~\rf{eq:deltalwkb} , and the Born approximation by Eq.~\rf{eq:born}.}
    \label{fig:delta2_hd}
\end{figure}

\begin{figure}[t]
    \centering
    \includegraphics[width=\columnwidth]{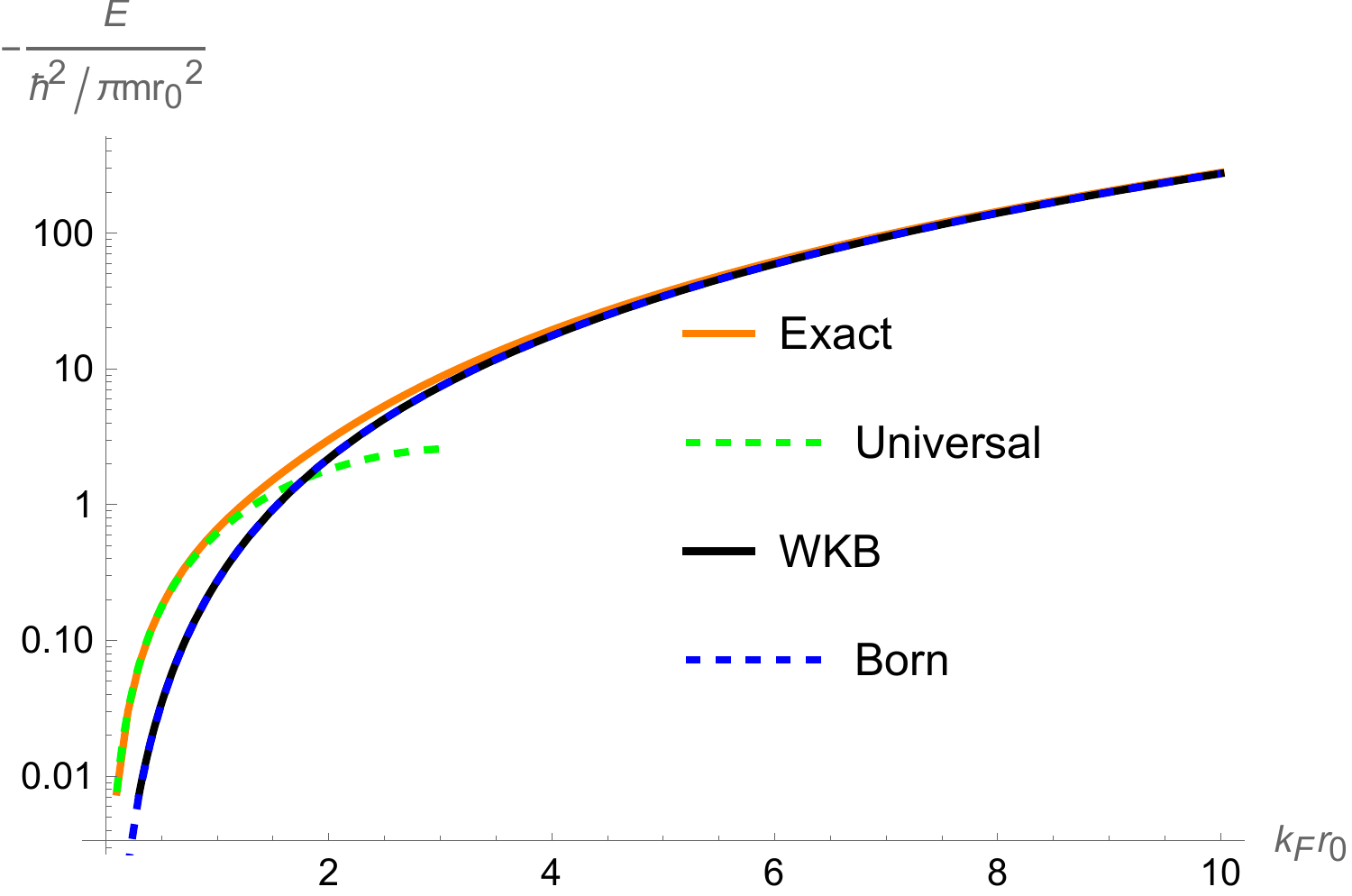}
    \caption{Energy $E$ of the square-well potential at unitarity $\gamma=\frac{\pi}{2}$ in the units of $\frac{\hbar^2}{\pi mr_0^2}$ computed from the exact expression for the phase shifts in Eq.~\rf{eq:swdeltalm}, universal result including the effective range correction Eq.~\rf{eq:Ewithreff}, asymptotic WKB result in Eq.~\rf{eq:WKBasymp},  and the Born approximation Eq.~\rf{eq:born}. }
    \label{fig:energy_wkb}
\end{figure}

\textit{General theory for $k_Fr_0\gg1$}. To get a more intuitive picture in the regime $k_Fr_0\gg1$, we first compute $\delta_l$ using the WKB method \cite{flugge2012practical}:
\begin{equation}
\label{eq:deltalwkb}
\begin{split}
&\delta_l(k) = \lim_{r'\to \infty}\left(\int_{r_t}^{r'}\sqrt{k^2-\frac{(l+\frac{1}{2})^2}{r^2}-\frac{2mV(r)}{\hbar^2}}dr\right. \\
&- \left.\int_{r_t^0}^{r'}\sqrt{k^2-\frac{(l+\frac{1}{2})^2}{r^2}}dr  \right), 
\end{split}
\end{equation}
where $r_t$ and $r_t^0$ are the corresponding turning points of the motion in the presence of the impurity potential and without it.  We used the Langer substitution $l(l+1)\to(l+1/2)^2$ to get the right expressions for the phase shifts \cite{flugge2012practical, landau2013quantum}. For simplicity, let us consider finite-range potentials that vanish beyond $r_0$. In this case, for a fixed value $k_Fr_0$, the relevant momenta are $k\leq k_F$. If for $k=k_F$, the turning point of the free motion occurs at $r_t^0>r_0$, the phase shifts are exactly zero. The condition when this occurs is:
$$r_t^0 = \frac{l+\frac{1}{2}}{k_Fr_0}r_0>r_0,$$
which implies that for $k_Fr_0>1$, we need to consider all angular momentum channels up to $l \sim k_Fr_0$. In practice, considering the contribution of one or two more angular momentum channels beyond $k_Fr_0$ is sufficient. For the square well potential, $\delta_l(k)$ can be computed analytically, but since this expression is not very illuminating, we do not reproduce it here. The behavior of the $\delta_l(k)$ for $l=2$ is presented in Fig.~\ref{fig:delta2_hd}. We observe that for a given value of $l$, the WKB result does not work below some small values of momenta ($k_\text{min}r_0 = l+1/2$). It turns out that in the regime $k_Fr_0\gg1$ the dominant contribution to the energy of the polaron comes from momenta close to $k_F$, so the fact that the WKB expression cannot be applied for very small $kr_0$ does not significantly affect the final result, as can be seen from Fig.~\ref{fig:energy_wkb}.

It is interesting to note that if $\gamma^2\ll k_Fr_0$, the phase shift for $k$ near $k_F$ are small and can be computed using the Born approximation \cite{landau2013quantum}:
\begin{equation}
\label{eq:born}
\sin\delta_l\approx \delta_l=-\frac{2mk}{\hbar^2}\int_0^{\infty}dr\,V(r)j_l^2(kr)r^2.
\end{equation}
The $l=2$ phase shift within the Born approximation for the unitary square well potential is plotted in  Fig.~\ref{fig:delta2_hd}. We observe again that the expression for the phase shifts computed within the Born approximation fails for small values of $k$, since the potential itself is not weak, yet in the regime $k_Fr_0\gg1$, their contribution to the polaron energy is negligible. Provided that $\gamma^2\ll k_Fr_0$, in the absence of the bound states, the energy of the polaron can be written as:
\begin{equation}
\label{eq:Elargedensity}
E = \frac{2}{\pi}\sum_{l}(2l+1)\int_{0}^{k_F}dk\,k^2\int_{0}^{\infty}dr\,r^2j_l^2(kr)V(r).    
\end{equation}
This formula can be applied to study general potentials. We just note that the condition of applicability of the Born approximation here is:
\begin{equation}
|V|\ll\frac{\hbar^2}{mr_0^2}(k_Fr_0),
\end{equation}
where $|V|$ is the characteristic value of the impurity potential. We note that if $|V|$ satisfies
\begin{equation}
|V|\ll\frac{\hbar^2}{mr_0^2},    
\end{equation}
then the Born approximation works for all $k$ \cite{landau2013quantum}, and the result of Eq.~\rf{eq:Elargedensity} can be used for arbitrary value of $k_Fr_0$. We observe that the energy of the polaron scales linearly with the strength of the potential in the above regime. For the case of the square well potential, using the analytical form of the WKB phase shifts, one can integrate over momenta and approximate the sum over $l$ by an integral to get the leading asymptotic result in powers of $k_Fr_0$:
\begin{equation}
\label{eq:WKBasymp}
E = -\frac{\hbar^2}{m\pi r_0^2}\left(\frac{\gamma^2}{9}(k_Fr_0)^3+\cdots\right).   
\end{equation}
Note that in obtaining the above result, we did not assume that $\gamma^2\ll k_Fr_0$, so it is valid as long as $k_Fr_0\gg1$. This means that if we consider values of $\gamma$, so that there are only a few bound states in the potential, each of which necessarily satisfies $|\varepsilon|< \gamma^2\hbar^2/(2mr_0^2)$, 
we observe that the contribution to the energy of the polaron from such bound states is negligible. As we start decreasing $k_Fr_0$ towards 1, we see that the bound states will start playing a more prominent role, and their presence needs to be included.

\textit{Application to Rydberg polarons} -- Finally, let us comment on how our results are relevant for the study of a Rydberg impurity that has gained a lot of attention in recent years \cite{schmidt2016mesoscopic, sous2020rydberg, gievers2024probing, durst2024phenomenology}. The microscopic impurity-fermion potential has the following form \cite{sous2020rydberg, massignan2025polarons}:
\begin{equation}
\label{eq:rydberg}
V_\text{Ryd} = \frac{2\pi \hbar^2}{m_e}|\psi_n(r)|^2, 
\end{equation}
where $a_e<0$ is the electron–ground-state-atom scattering length, $m_e$ is the mass of the electron, and $\psi_n(r)$ is the Rydberg atom wavefunction. The characteristic size of the potential $r_0$ scales with $n$ as $r_0\propto n^2$ \cite{gallagher2006rydberg}, so the regime of $k_Fr_0\gg1$ can be achieved experimentally, and it is necessary to include all angular momentum channels up to $l\sim k_Fr_0$ in the calculations. At the same time, such potential supports multiple bound states \cite{massignan2025polarons}, so their contribution to the energy needs to be included as well. Their exact values can be computed numerically by solving the Schr\"odinger equation with the potential in Eq.~\rf{eq:rydberg}. The general behavior should be in accordance with the results presented here, and the effect of bound states and the higher angular momentum channels should be visible in the spectroscopic signatures, such as the impurity's absorption spectrum \cite{knap2012time}.

\textit{Summary and outlooks} -- We applied Fumi's theorem to study the properties of static Fermi polarons with finite-range impurity-fermion interaction. We showed how this formalism allows one to study polarons beyond the $s$-wave scattering regime in a systematic way. This allows one to extract all the essential ingredients that are relevant for the understanding of immobile polarons with finite-range interactions, such as ionic and Rydberg impurities. A natural extension of our work would be a generalization to finite impurity masses in the spirit of \cite{chen2025massgapdescriptionheavyimpurities}.

\vspace{5mm}
\begin{acknowledgments}
We acknowledge P. Massignan, G. Astrakharchik, A. Duspayev, E. Zhao, and W. V. Liu for insightful discussions and useful comments on the manuscript. This work was supported by AFOSR Grant No.~FA9550-23-1–0598.
\end{acknowledgments}

\bibliography{UnitaryPolaron}

\end{document}